\documentclass[prb,aps,amsmath,preprint,]{revtex4-1}
\usepackage{graphicx}

\usepackage{dcolumn}

\usepackage{bm}
\usepackage{subfigure}

\begin{document}

\title{Effect of van der Waals interactions on the structural and elastic properties of black phosphorus}

\author{S. Appalakondaiah and G. Vaitheeswaran$^*$}
\affiliation{Advanced Centre of Research in High Energy Materials (ACRHEM),\\
University of Hyderabad, Prof. C. R. Rao Road, Gachibowli, Andhra Pradesh, Hyderabad- 500 046, India.}

\author{S. Leb\`egue}
\affiliation{Laboratoire de Cristallographie, R\'esonance Magn\'etique et Mod\'elisations (CRM2, UMR CNRS 7036),Institut Jean Barriol, Nancy Universit\'e BP 239,\\
 Boulevard des Aiguillettes, 54506 Vandoeuvre-l\`es-Nancy, France.}

\author{N. E. Christensen and A. Svane}
\affiliation{Department of Physics and Astronomy,\\
Aarhus University, DK-8000 Aarhus C, Denmark.}

\date{\today}

\begin{abstract}
The structural and elastic properties of orthorhombic black phosphorus have been investigated using first-principles calculations based on density functional theory.
The structural parameters have been calculated using the local density approximation (LDA), the generalized gradient approximation (GGA), and
 with several dispersion corrections to include van der Waals interactions. It is found that the  dispersion corrections improve
 the lattice parameters over LDA and GGA in comparison
 with experimental results. The calculations
 reproduce well the experimental trends under pressure and show that van der Waals interactions are most important for the crystallographic $b$-axis,
 in the sense that they have the largest effect on the bonding between the phosphorus layers. The
 elastic constants are calculated and are found to be in good agreement with experimental values. The
 calculated C$_{22}$ elastic constant is significantly larger than the C$_{11}$ and C$_{33}$ parameters,
 implying that black phosphorus is stiffer against strain along the $a$-axis than along the $b$- and $c$-axes. From the calculated elastic constants,
 the mechanical properties such as bulk modulus, shear modulus, Young's modulus and Poisson's ratio are obtained. The calculated Raman active optical phonon frequencies
 and their pressure variations are in excellent agreement with available experimental results.
\end{abstract}

\maketitle
\section{INTRODUCTION}
Phosphorus, an extraordinary element in the group 5A  of the Periodic Table, exists in different allotropic modifications including white, black and amorphous red at ambient conditions.\cite{Boocker}
Among these allotropic forms black phosphorus is the most stable one.\cite{Morita} It is a semiconductor\cite{Akahama} that crystallizes in an orthorhombic structure.\cite{Cartz} At ambient conditions, orthorhombic black
phosphorus has a puckered layered structure where each phosphorus atom is covalently bonded with three neighboring phosphorus atoms in the same layer, the layers being weakly coupled by van der
Waals (vdW) forces.
At high pressures  orthorhombic black phosphorus transforms to a rhombohedral structure around 5 GPa, and transforms further on to a simple cubic structure
at 10 GPa.\cite{John,Duggin,Kikegawa,Suzuki,Okajima} At very high pressures of 137 GPa and 262 GPa, further structural phase transitions have been studied by Akahama et al.\cite{Akahama137,Akahama262} using x-ray diffraction.
The measurements of the physical properties of black phosphorus such as the anisotropic electrical resistivity\cite{Akahama} and the optical
reflectance  have been reported,\cite{Maruyama, Narita} as have been angle resolved ultraviolet photoemission spectra,\cite{Takahashi} infrared absorption spectra,\cite{Ikezawa} and
phonon dispersions obtained by inelastic
neutron scattering.\cite{Fujii}

On the theoretical side, investigations of the electronic structure\cite{Erik, Rajagopal, Rajeev2003} using the linear muffin tin orbital (LMTO)
method,\cite{OKA} and optical properties using the self consistent pseudopotential method were published.\cite{Morita,Asahina} Lattice dynamical properties of black phosphorus including
phonon dispersion relations were obtained\cite{Kaneta} using a valence force field method. A number of studies discuss the high-pressure behavior of black phosphorus, notably the structural stability and
the electronic
structure.\cite{Rajeev2003,Schiferl,Chang,Shirotani,Kawamura,Shindo,Ehlers,Prytz,Clark,Grin}
Recently,  Du et al.\cite{Yanlan} performed calculations for single layered and bulk black phosphorus and reported that
black phosphorus may be used as anode material for lithium ion batteries.

Since black phosphorus is a layered structure material that binds through van der Waals interactions between the layers it is necessary to include the vdW
 interactions
in the theoretical calculations of the ground state structural properties of this system.
 Van der Waals forces, or dispersive forces, result
 from the interaction between 
 fluctuating multipoles without the need of an overlap of the electron densities.
 The standard density functional theory (DFT) functionals such as the
 local density approximation (LDA), the generalized gradient approximation (GGA), and various hybrid functionals do not describe properly the vdW forces, which are non-local in nature.
 To overcome this difficulty, we have used dispersion corrected density-functional theory (DFT-D) approaches,
 which enhance the total-energy functional by  a damped\cite{Grimme} interatomic pairwise potential energy of the form C$_6$R$^{-6}$,
 where $R$ denotes the interatomic distance.
The  main focus of the present work is to calculate the ground state structural and dynamical properties of black phosphorus, and to compare several   schemes for inclusion of vdW interactions.

 The paper is organized as follows: The next section  presents the computational methods used, and the results and discussions are presented
in section III. The last section, IV, contains a summary.

\section{COMPUTATIONAL DETAILS}
 The calculations were performed using the CAmbridge Series of Total Energy Package (CASTEP)\cite{Payne, Segall}
 code based on DFT using Vanderbilt-type
 ultra soft pseudo potentials\cite{Vanderbilt} for the electron-ion interactions.
 The exchange-correlational potential of Ceperley and Alder\cite{Ceperley} as parameterized by
 Perdew and Zunger\cite{PPerdew}(CAPZ) in LDA, and the GGA schemes of Perdew and Wang\cite{Wang} (PW91) and of Perdew-Burke-Ernzerhof\cite{Perdew} (PBE)
 were used.  Different DFT-D approaches to treat vdW interactions were employed, notably the Ortmann, Bechstedt and Schmidt\cite{OBS} (OBS) correction to PW91,
 as well as the Tkatchenko and Scheffler\cite{TS} (TS) and
 Grimme\cite{Grimme}  corrections to PBE.
 In the three methods, a correction of the form $ E_{ij} = f(R_{ij}) \times C_6^{ij} \times R_{ij}^{-6} $ is added to the DFT total energy for each pair $(ij)$
  of atoms separated by a distance $R_{ij}$. $f(R)$ is a damping function which is necessary to avoid divergence for small values of $R$, and
  $C_6^{ij}$ is the dispersion coefficient for the atom pair $(ij)$. The three methods use a slightly different damping function
  (see Refs. \onlinecite{OBS}, \onlinecite{Grimme} and \onlinecite{TS}
   for details), but while the $C_6^{ij}$ coefficients in the Grimme and OBS methods depend only on the chemical species, the TS method calculate them
    for every structure by using the fact that the polarisability depends on the volume (see Ref. \onlinecite{TS}). In the present study, we did not attempt
     to tweak the various coefficients of the different methods and kept them as in the original publications. In particular, in black phosphorus enters only
     one $C_6$ coefficient, for the P-P interaction, with values of
      $C_6= 103.6$, $81.26$, and $110.5$ eV$\cdot$\AA$^6$ in the OBS, Grimme and TS (unscaled) approaches, respectively.
      These methods have shown to be realiable and are often used for electronic structure calculations.
For instance, Grimme's method has been used recently to study the binding properties of various solids\cite{sebvasp}
 and the structure and dynamics of molecular spin crossover compounds.\cite{sebpccp} The OBS method is also widely used, as exemplified 
 by the study\cite{obsapp1} of the interaction between NO$_2$ molecules and the Au-111 surface. The method of Tkatchenko
 and Scheffler is also very popular and was used recently to study the hydrogen bonds of ice under pressure\cite{ts1app}
 and the interlayer sliding energy landscape of hexagonal boron nitride.\cite{ts2app}

 To perform our calculations, a plane wave  cut-off energy of 520 eV is used and the first
 Brillouin zone  of the unit cell is sampled according to the Monkhorst-Pack scheme,\cite{Monkhorst}
 by means of a (9x6x8) k-point sampling. The Pulay\cite{Pulay} density mixing
 scheme was applied for the electron energy minimization process.
 During the structural optimization process iterations were continued until 
 the change in total energy  was  less than $5\cdot 10^{-6}$ eV/atom and the
  maximum displacement of atoms less than $5\cdot 10^{-5}$ \AA.
 The elastic constants are calculated using the volume-conserving strain technique as implemented in the CASTEP code.\cite{Mehl, Ravindran}

At ambient conditions black phosphorus crystallizes  in a base centered orthorhombic structure (see Fig. 1) with the space group of C\emph{mca} (no. 64). The experimental
lattice parameters\cite{Cartz} are $a=3.31$ \AA, $b=10.47$ \AA\ and $c=4.37$ \AA. The crystal structure   consists of puckered layers of P atoms with three short covalent bonds (two at 2.22 \AA, and one at 2.28 \AA), with  vdW bonding along the $b$-axis,  the vertical distance between layers being 3.27 \AA.
The orthorhombic unit cell contains 8
 phosphorus atoms in the crystallographic ($8f$) positions\cite{Cartz} $(0, u, v)$. The nearest and next-nearest atomic separations between the layers ($d_1$ and $d_2$ in Fig. 1) are 3.6 \AA\ and 3.8 \AA.

In the orthorhombic crystal structure there are nine independent elastic constants\cite{Ravindran}: C$_{11}$, C$_{22}$, C$_{33}$,  C$_{44}$, C$_{55}$,  C$_{66}$, C$_{12}$, C$_{13}$, C$_{23}$.
Here, we follow the conventions of the experimental works\cite{Hanayama,Masahito} in which the   indices $1,2,3$ are defined
to correspond to the crystallographic directions $X,Y,Z$ ($ c,a,b $), respectively, as indicated in Fig. 1.
To calculate the elastic constants, the equilibrium crystal structure is deformed by the appropriate strain
and the variation in total energy gives the elastic constants. This is explained in detail for orthorhombic crystals in  Ref. \onlinecite{Ravindran}. The phosphorus internal coordinates $u$ and $v$ are freely varied during these deformations. 

\section{Results and discussions}
 \subsection{Structural properties}

 The theoretical equilibrium
  crystal structure was obtained by  full structural optimization including lattice constants and the internal coordinates $u$ and $v$, using LDA, GGA, and
  the three considered versions of GGA-D.
  The calculated
    structural parameters  are presented in Table I.
   Taking the experimental ambient volume as reference, we find the difference between calculated and experimental volume of -11 \% with LDA, +8 \% with PW91, and  +10 \% with PBE.
   The large errors reflect that none of these approximations to the exchange-correlation potential are able to capture
  correctly the nature of the interactions in black phosphorus. The DFT-D schemes  perform much better
   (see  Table I), the errors in predicted equilibrium volumes being only  +1.5 \%, +2.9 \%, and -0.3 \%, respectively, for the PW91-OBS, PBE-TS, and PBE-Grimme parametrizations.
   Considering the three lattice parameters, it is clear that the
   largest error is found for the $b$-parameter, the characteristic dimension in the Z-direction, i.e. perpendicular to the 
   covalent P layers. Thus, the interlayer interaction is dominated by dispersive forces (vdW),
   and an accurate description of this type of interaction is important. Comparison of the data for GGA-PBE and GGA-PBE-Grimme in Table I shows that the inclusion of  vdW interaction reduces the $c$-parameter by $\sim 3$ \%, while the
$b$-parameter is reduced by $\sim 7$ \%.
   On the other hand, the internal parameters $u$ and $v$ are about equally well
   predicted by all functionals considered, probably because these parameters are determined by the P-P covalent bonds, which are usually well described by LDA and GGA.
     The overall best agreement with experimental parameters is achieved with the PBE-Grimme functional.
    Our LDA calculations are in reasonable agreement with the LDA calculations performed by Ahuja\cite{Rajeev2003} (see Table I).
    The minor differences are most likely due to the different parametrizations of the LDA functional used and to the different basis sets
     (plane waves in the present work versus full potential LMTO in Ref. \onlinecite{Rajeev2003}).


\begin{figure}
\centering
\includegraphics[width=\linewidth,clip]{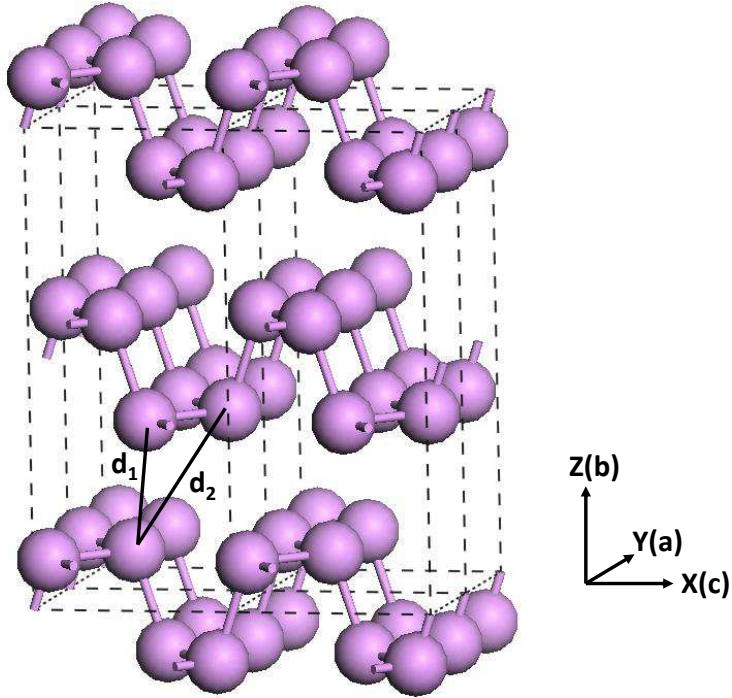}
\caption{(Color online) Crystal structure of black phosphorus. The nearest and next-nearest interlayer atomic distances are shown ($d_1$ and $d_2$).}\label{Fig 1}
\end{figure}

\begin{table}[tb]
\caption{The calculated ground state properties of orthorhombic black phosphorus at ambient pressure.
$a$, $b$ and $c$ are the lattice parameters, $V$ the volume of the orthorhombic unit cell, and $u$ and $v$ the internal crystallographic parameters.
CAPZ denotes the LDA parametrization  of Refs. \onlinecite{Ceperley,PPerdew}, PW91 and PBE the
GGA parametrization of Refs. \onlinecite{Wang,Perdew}, respectively. For the van der Waals parametrization, the parametrizations of Ref. \onlinecite{OBS} (OBS), Ref. \onlinecite{TS} (TS) and
Ref. \onlinecite{Grimme} (Grimme) are considered.
Also included are the results of the LDA calculations by Ahuja\cite{Rajeev2003} with the Hedin-Lundqvist parametrization (LDA-HL)\cite{HL}.
 }
\begin{ruledtabular}
\begin{tabular}{cccccccccccccccccc}
 Method &XC& a(\AA) & b(\AA) & c(\AA) & V(\AA$^3$) & u & v&   \\ \hline
 DFT    &  LDA-CAPZ & 3.28 & 10.10 & 4.07 & 134.9 & 0.1059 & 0.0711\\
        & LDA-HL$^a$& 3.24 & 10.19 & 4.24 & 140.0 & 0.1013 & 0.0789 \\
        & GGA-PW91  & 3.29 & 11.07 & 4.52 & 164.5 & 0.0950 & 0.0871\\
        & GGA-PBE   & 3.28 & 11.22 & 4.54 & 167.1 & 0.0935 & 0.0876\\
 \hline
DFT-D   & GGA-PW91-OBS   & 3.29 & 10.63 & 4.41 & 154.1  & 0.0996 & 0.0832\\
        & GGA-PBE-TS     & 3.29 & 10.82 & 4.39 & 156.3  & 0.0979 & 0.0830\\
        & GGA-PBE-Grimme & 3.30 & 10.43 & 4.40 & 151.3  & 0.1017 & 0.0833\\
\hline
Exp$^b$ &                & 3.3133 & 10.473 & 4.374 & 151.77 & 0.1034 & 0.0806\\
 \end{tabular}
\end{ruledtabular}
$a:$ Ref. \onlinecite{Rajeev2003}. \\
$b:$ Ref. \onlinecite{Cartz}.
\end{table}

\begin{figure}[h!]

\includegraphics[width=4in,clip]{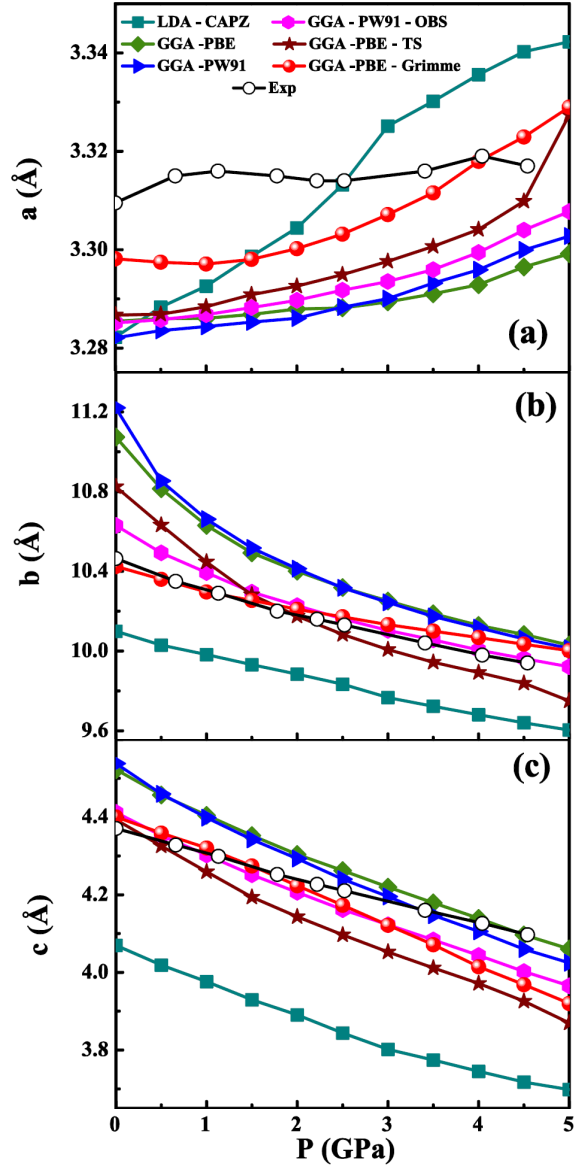}
\caption{(Color online) (a), (b) and (c): pressure dependence of lattice parameters $a$, $b$ and $c$ up to 5 GPa  within LDA, GGA and GGA-D, compared to experimental values (Ref. \onlinecite{Akai}).
Note the different scales on the vertical axes, in particular the small scale in (a).}\label{Fig_abc}
 \end{figure}

To investigate the response of black phosphorus to external pressure, hydrostatic compression was applied to the unit cell in the pressure range
 of 0 to 5 GPa in steps of 0.5 GPa. This was done by variable cell optimization under the constraint of a diagonal stress tensor with  values of the diagonal elements fixed   to specify the
 desired pressure. The change of the lattice parameters $a$, $b$ and $c$ with pressure is compared with the experimental values\cite{Akai} and shown in Fig. \ref{Fig_abc}.
 It is seen that the $a$ lattice parameter shows only a weak variation under pressure, which reflects the strong bonding along this direction. This fact is correctly
  captured by all the functionals that we have used, although we obtain a larger variation than in the experiments. In fact all functionals predict a slight increase in $a$ with pressure.
  The variation is much larger for $b$ and $c$.
  For the $b$ parameter, a reduction of $\sim 0.52$ \AA\ is observed experimentally over the pressure range of 0-4.5 GPa.\cite{Akai}  This is correctly reproduced when the vdW
    correction is included, but it is about twice as large with the pure PW91 and PBE functionals. As for the $c$ parameter, experimentally a
  reduction by approximately 0.27 \AA\ is seen in this pressure range (Fig. 2(c)). This variation is overestimated by all   methods that we have tested, even those corrected
      for vdW interactions.
 
These results are also quantified in terms of the  pressure coefficients defined as (with $x$ being either $a$, $b$, or $c$)
\begin{equation}
	\label{gamma}
	\gamma(x)=\left. \frac{dx}{dP}\right|_{P=0}
\end{equation}
 which  are shown in Table \ref{tab:presder}. The pressure  coefficient   for the $a$ lattice parameter is very
 small for all approximations, in accordance with the experimental data.\cite{Cartz}
 For the $b$ lattice parameter, the pressure coefficient varies significantly between the different approximations, but comes very close to the experimental value with
 Grimme's corrected functional. The magnitude of the pressure coefficient for the $c$ parameter is roughly twice as large as found experimentally, with
  the best agreement found again with Grimme's method.

\begin{figure}[h!]
\includegraphics[width=\linewidth,clip]{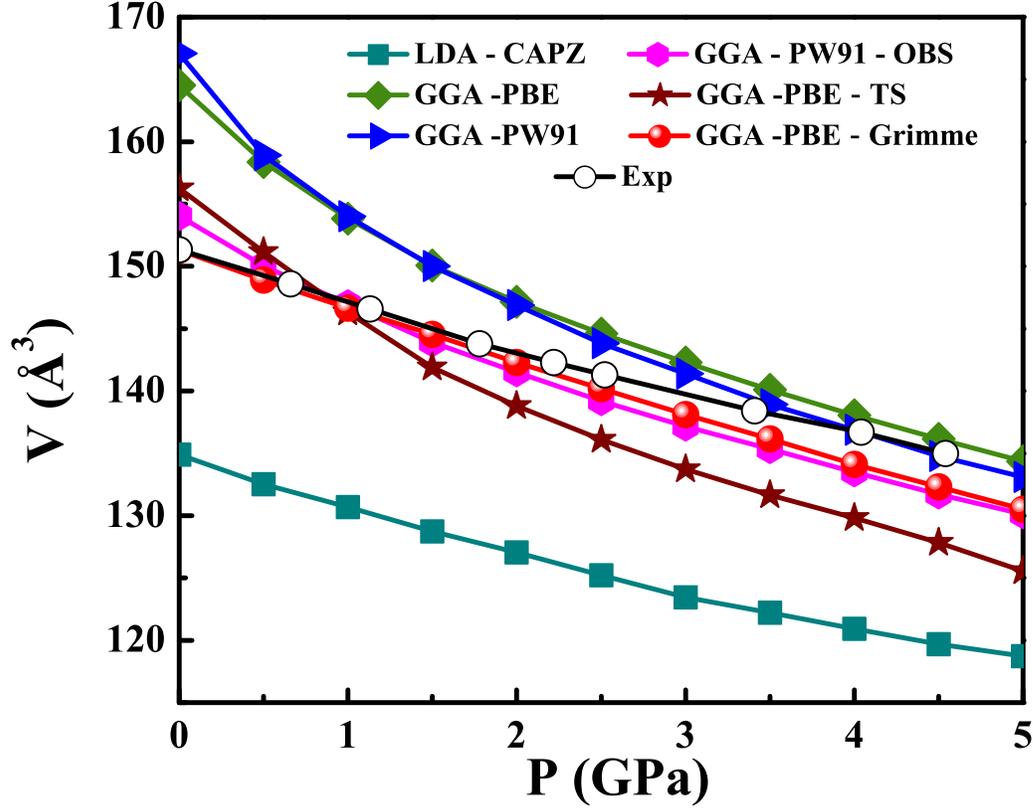}
\caption{(Color online) Comparison of the pressure dependence of the crystal volume as calculated  within LDA, GGA and GGA-D, compared to experimental volumes (Ref. \onlinecite{Akai}).}
\label{Fig_vol}
 \end{figure}

 The calculated pressure-volume (PV) relation of black phosphorus is shown in Fig. \ref{Fig_vol} and compared to experimental results.\cite{Akai} Apart from the offset in equilibrium volumes discussed above, it is
 noticable that both the GGA and the GGA-D calculations find black phosphorus more compressible than observed experimentally.
 
The  slopes of the calculated $V(P)$ relations at $P=0$ are larger than seen in experiments,\cite{Akai} with the exception of the Grimme approach. 
 This implies that the bulk moduli calculated with these approaches are significantly smaller than the experimental value. For the Grimme functional
the initial slope is very similar to the experimental value, and the calculated bulk modulus within the Grimme approach is $B=31.3$ GPa, 
which is in good agreement with experimental values of 32.5 GPa\cite{Cartz}
and 36 GPa.\cite{Akai} Therefore, comparing all the theoretical approaches applied here for black phosphorus, Grimme's functional shows overall best performance, and it is noticeable that the dispersion correction provided by the Grimme functional has the right magnitude at low pressure, but seems to be slightly too strong for elevated pressures.  However, one should bear in mind
 that the experiments under pressure are performed at room temperature, while  the theory pertains to $T=0$.

\begin{figure}[h!]
\includegraphics[width=4in,clip]{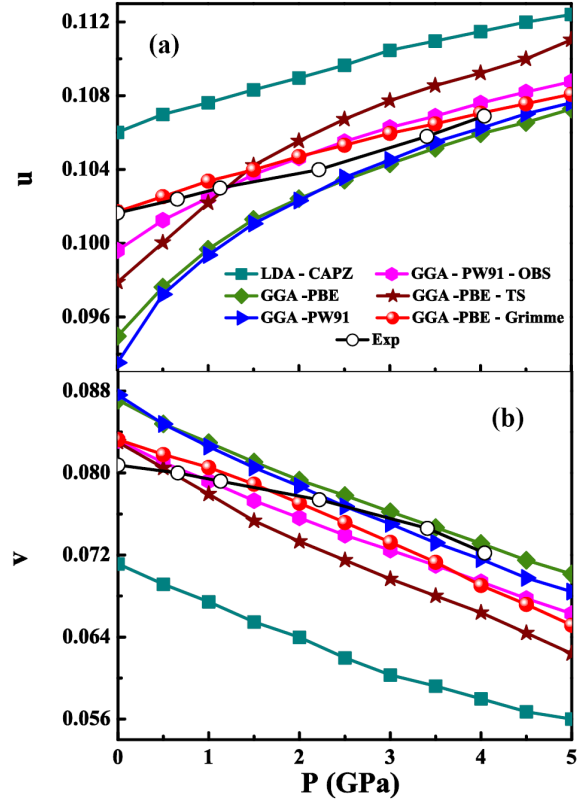}
\caption{(Color online)   Internal coordinates,  $u$ (a) and $v$ (b), of black phosphorus up to 5 GPa as calculated within LDA, GGA and GGA-D, compared to experimental values (Ref. \onlinecite{Akai}).}
\label{Fig_uv}
\end{figure}
\begin{figure}[h!]
\includegraphics[width=4in,clip]{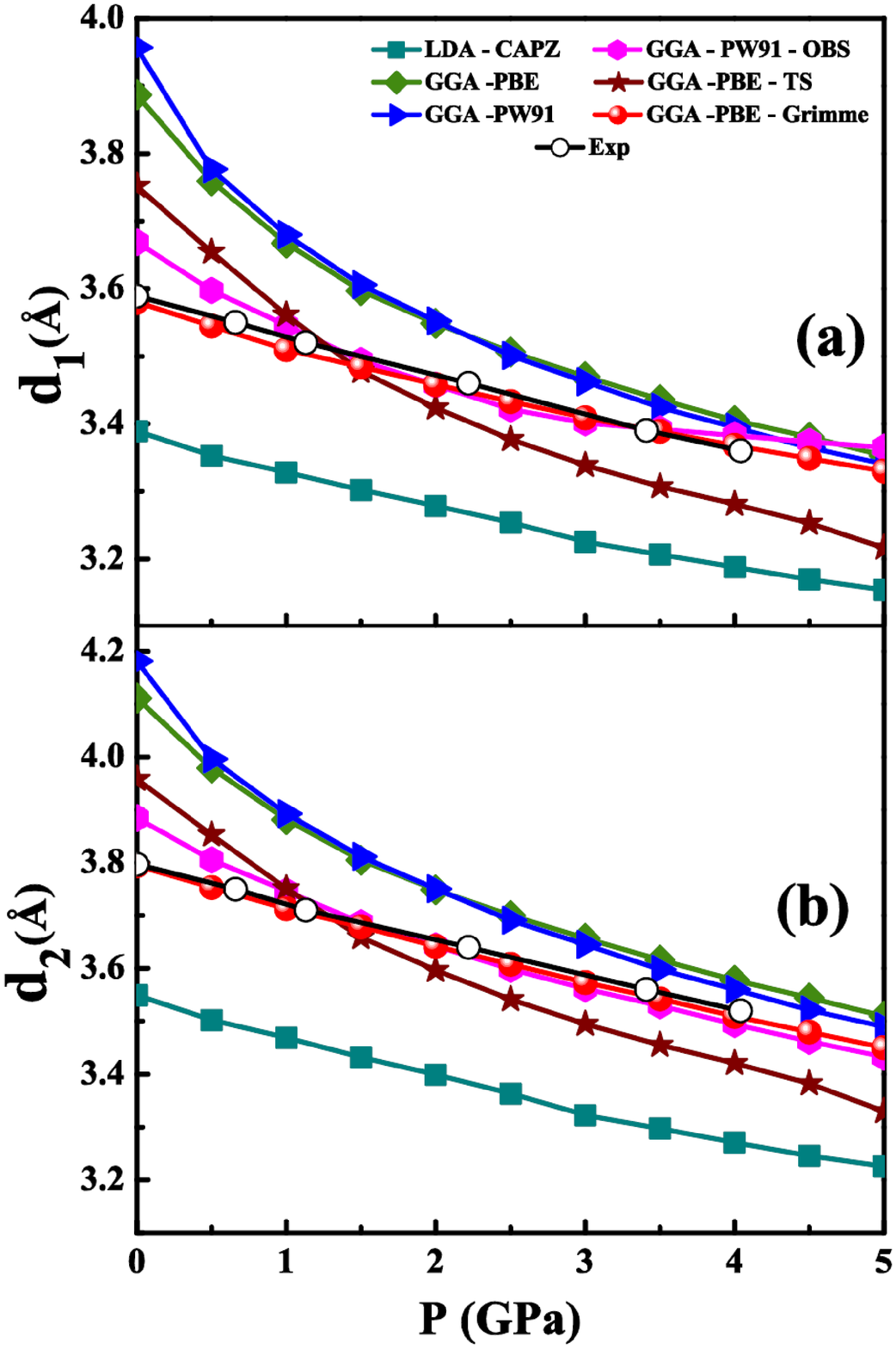}
\caption{(Color online) Calculated (a): nearest neighbour ($d_1$) and   (b): next-nearest nieghbor $(d_2$) interlayer atomic separations (cf. Fig. 1) of black phosphorus up to 5 GPa with and without   dispersion correction and compared with experiments (Ref. \onlinecite{Akai}).}
\label{Fig_d1d2}
\end{figure}

The variation under pressure of the internal coordinates $u$ and $v$ are illustrated in Fig. \ref{Fig_uv}. All functionals capture the increasing (decreasing) trends of $u(P)$ ($v(P)$)
 found in the experiment.\cite{Akai} Clearly, the LDA shows the poorest agreement with the experimental data, and the vdW corrected GGA show marginally better agreement with experiments than the pure GGA functionals.
 Also, Fig. \ref{Fig_d1d2} illustrates the pressure dependance of the nearest and next nearest interlayer atomic distances, $d_1$ and $d_2$ in Fig. \ref{Fig 1}. It is seen that LDA grossly underestimates
 and the pure GGA functionals overestimate these parameters. For the vdW corrected functionals, the agreement with experiments is much better,
 and the pressure evolution of the inter-layer atomic distances is best reproduced with the Grimme functional. This is of course in line with the better
  agreement found for the equilibrium volume and the other properties under pressure.


\begin{table}[tb]
\caption{The calculated first order pressure coefficients of the lattice parameters (Eq. (\ref{gamma})) of orthorhombic black phosphorus, using LDA-CAPZ as well as GGA
with and without dispersive corrections (see caption of Table I). The experimental results are taken from Ref. \onlinecite{Cartz}. Units are $10^{-3}$\AA$\cdot$GPa$^{-1}$.
}
\begin{ruledtabular}
\begin{tabular}{cccccccccccccccccc}
          && DFT &&            &  DFT-D &&& Expt. (Ref. \onlinecite{Cartz}) \\\hline
 axis     &LDA  &PW91    & PBE  &   &  PW91-OBS & PBE-TS   & PBE-Grimme &                 \\ \hline
 a        &11   &$\sim 1$ &  2  &   &   3       &  2       &  $\sim$ 0       &  $\sim 0$       \\
 b        &-123 & -460   & -566 &   &  -261     &  -411    &   -124     &  -125           \\
 c        &-101 & -125   & -135 &   &  -118      &  -147    &   -82      &  -55            \\
 \end{tabular}
\end{ruledtabular}
\label{tab:presder}
\end{table}

\subsection{Elastic properties}
Elastic constants are  fundamental mechanical parameters for crystalline materials and   describe  the stiffness of the material against  externally applied strains.
In the present study, we have calculated elastic properties for single crystal as well as polycrystalline black phosphorus within the GGA and vdW corrected GGA approximations.
 The calculated elastic constants are presented in Table III. All the calculated elastic constants satisfy   Born's mechanical stability criteria\cite{Born} implying that the system is in a mechanical stable state. The elastic constants have been determined for black phosphorus in Refs. \onlinecite{Hanayama,Masahito} by ultrasound velocity measurements.
 Following the convention of these works,  the C$_{11}$, C$_{22}$, and C$_{33}$ elastic constants   directly relate  to sound propagation along the crystallographic $c$, $a$, and $b$-axes, respectively,
 and reflect the stiffness to uniaxial strains along these directions.\cite{Ravindran} The reported experimental values show some scatter, which
 probably reflects the difficulty of the measurements. Also, the elastic constants show some variation with temperature, as discussed in Ref. \onlinecite{Masahito}.

 From  Table III, it is seen that the calculated elastic constants show some variations when evaluated with different functionals,
 which is a combined effect of the different effective forces and the different equilibrium structures obtained with the various functionals.
 The   C$_{22}$ elastic constant  is significantly larger than   C$_{11}$ and C$_{33}$.
 This shows that the material is stiffer for strains along the $a$-axis than along the $b$- and $c$-axes, consistent with the results shown in  Fig. \ref{Fig_abc}, which again
  can be attributed to the strong covalent bonds between P atoms  along the crystallographic $a$-axis.  For C$_{22}$, the calculated values are consistent between the different functionals and in excellent agreement with the experimental value of Ref. \onlinecite{Hanayama}, while the experimental value of  Ref. \onlinecite{Masahito} is significantly larger.
The C$_{11}$ is relatively constant among the considered functionals and again in fair agreement with the experiment of Ref. \onlinecite{Hanayama}, while the value obtained in
Ref. \onlinecite{Masahito} is  larger.
The C$_{33}$ describes stiffness to stress along the $b$-axis, and this has a small value when evaluated in pure GGA, while the vdW-corrected methods give larger values, for the OBS and TS
flavors in reasonable agreement with the experimental values, while the Grimme approach significantly overestimates this parameter. The C$_{44}$ and C$_{55}$ constants are relatively small and in good agreement between theory and experiment (although the value of C$_{55}$ calculated with PBE-TS stands out as too large).
On the other hand, for C$_{66}$ the calculated values are fairly consistent and in good agreement with the experimental value of Ref. \onlinecite{Masahito}, while  the measured
value reported by Ref. \onlinecite{Hanayama} is a factor 3-4 smaller. Among the off-diagonal elastic constants, C$_{12}$ is large and relatively consistent between the various functionals, while the
C$_{13}$ and C$_{23}$ are small, and in some cases take negative values. There are no reported experimental values of these quantities.
From the single crystal elastic constants  the single-crystal bulk modulus B
may be found.\cite{Ravindran}  This gives a value of $B=30.7$ GPa (using the Grimme functional), which is in good agreement with the experimental values of $B=32.5$ GPa\cite{Cartz}
or $B=36$ GPa,\cite{Akai} as well as with the value derived from the $V(P)$ relation in Fig. \ref{Fig_vol}, $B=31.3$ GPa (also in the Grimme approximation).
Altogether, the agreement between the measured and the calculated elastic constants is fair while not perfect.

\par The  bulk and shear moduli of polycrystalline black phosphorus may be obtained from the elastic constants using the Voigt-Reuss-Hill approximations. \cite{Voigt, Reuss, Hill} Here, the Voigt and Reuss
approximations represent extreme values, and Hill recommended that their average is used in practice for polycrystalline samples. Their definitions for orthorhombic systems are given by Ref. \onlinecite{Ravindran} and their values for black phosphorus are quoted in Table IV.
Using the calculated bulk and shear moduli in the Hill approximation, $B_H$ and $G_H$,  we have  evaluated the Young's modulus, $E$, and Poisson's ratio, $\sigma$. Young's modulus
 reflects the resistance of a material towards uniaxial tensions, while Poisson's ratio gives information about the stability against shear strain.
 The results obtained are summarized in Table IV. It is seen that for polycrystalline black phosphorus,  $B_H  >   G _H$, which implies that the parameter limiting the mechanical stability of black phosphorus is the shear modulus  $G_H$. The  bulk   and shear moduli provides  information about the brittle-ductile nature of materials through Pugh's ratio. \cite{Pugh} According to Pugh, a $B_H/ G_H$ ratio  below (above)  1.75   indicates the \textit{brittle} (\textit{ductile}) nature of the material. For black phosphorus the  calculated  value  is  $B_H/ G_H=1.31$, i. e.  indicating a  brittle nature of the material. Generally, Poisson's ratio    is connected with the volume change of a material during uniaxial deformation and the nature of interatomic forces. If $\sigma$ is 0.5, no volume change occurs, whereas lower than 0.5 means that large volume change is associated with elastic deformation.\cite{Ravindran} In our calculation $\sigma=0.30$ is obtained, which shows that a considerable volume change can be associated with the deformation in
black phosphorus.

 The anisotropy in bonding between atoms in different planes is related to an anisotropic thermal expansion.
 This anisotropy  plays an  important role for all orthorhombic systems, as they influence the
 durability of the material to repeated heating cycles. The shear anisotropy factors measure this effect in terms of the elastic constants (see definitions for orthorhombic crystals in Ref. \onlinecite{Ravindran,bheem}).
 If the anisotropy factor is unity it represents complete elastic isotropy, while values smaller or greater than unity measure the degree of elastic anisotropy.
 For black phosphorus the calculated shear anisotropy factors A$_1$ is close to unity, which shows that the resistance to
 shears of $\langle 100 \rangle$ planes (i.e. planes having $c$ as a normal) along the $a$ and $b$ directions are approximately equally. In contrast, the A$_2$ and   A$_3$ anisotropy factors are very different from unity,
 implying that shears of planes orthogonal to $a$ and of planes orthogonal to $b$ both behave very anisotropically.


\begin{table}[tb]
\caption{
Single-crystal elastic constants (C$_{ij}$, in GPa) and bulk modulus ($B$ in GPa) of black phosphorus. All quantities are calculated at the
respective theoretical equilibrium volumes using the various GGA and GGA-D functionals.}
\begin{ruledtabular}
\begin{tabular}{cccccccccccccccccccccccc}
          & DFT &&            &  DFT-D &&& Expt. \\\hline
 parameter  &PW91    & PBE &    &  PW91-OBS & PBE-TS   & PBE-Grimme& Exp & \\ \hline
C$_{11}$    &45.2  &40.7&  &44.4  & 36.8& 52.3 & 55.1$^a$,   80$^b$ \\
C$_{22}$    &186.7 &183.4& &192.4 &185.9&191.9 & 178.6$^a$, 284$^b$ \\
C$_{33}$    &19.2  &13.0&  &40.5  &30.6 &73.0  & 53.6$^a$,   57$^b$ \\
C$_{44}$    &10.3  &8.6&   &15.9  &31.2 &25.5  & 21.3$^a$, 11.1$^a$, 17.2$^b$ \\
C$_{55}$    &4.2   &2.6&   &5.6   &23.4 &8.8   & 5.5$^a$,  10.8$^b$ \\
C$_{66}$    &48.3  &48.5&  &57.3  &57.4 &63.6  & 14.5$^a$, 15.6$^a$, 59.4$^b$ \\
C$_{12}$    &32.3  &30.0&  &35.6  &31.5 &40.8  &--- \\
C$_{13}$    &-4.3  &-4.6&  &-5.3  &-0.9   &0.6  &  --- \\
C$_{23}$    &1.0  &-1.6&   &1.5   &-0.6  &8.3  & --- \\
B           &11.6 &8.1 &   &18.6  &16.2 &30.7  & 32.5$^c$, 36$^d$ \\
\end{tabular}
\end{ruledtabular}
a: Ref. \onlinecite{Hanayama} (room temperature); b: As read from figures of Ref. \onlinecite{Masahito} (T=0 K); c: Ref. \onlinecite{Cartz}; d: From $PV$-data of Ref. \onlinecite{Akai}.
\end{table}

\begin{table}[tb]
\caption{Calculated values of the polycrystalline   bulk ($B_X$) and shear moduli ($G_X$) in the Voigt, Reuss and Hill approximations ($X=V, R, H$, respectively). Units are GPa.
Also listed are the Young's modulus, $E$ (in GPa),  Poisson's ratio, $\sigma$, and the shear anisotropy factors, A$_i$. All values are calculated at the theoretical equilibrium volume using the
PBE-Grimme functional.}
\begin{ruledtabular}
\begin{tabular}{ccccccccccccc}
Parameter & B$_V$ & B$_R$ & B$_H$ & G$_V$ & G$_R$ & G$_H$ & E    & $\sigma$ & A$_1$ & A$_2$ & A$_3$\\ \hline

           & 46.3  & 30.7  & 38.5  & 37.4  & 21.4  & 29.4  & 70.3 & 0.30     & 0.8    & 0.1   & 1.6\\

\end{tabular}
\end{ruledtabular}
\end{table}

%

\subsection{Transition to the $A$7 structure}



Black phosphorus transforms under pressure from the orthorhombic structure ($A$17) to a rhombohedral structure ($A$7)
 at roughly $P_t=5$ GPa.\cite{John,Duggin,Kikegawa,Suzuki,Okajima}
 This transition was studied previously by Schiferl\cite{Schiferl} using an empirical local pseudopotential, and was analysed by Burdett et al.\cite{Burdett}
  using the concept of orbital symmetry conservation.   Chang and Cohen\cite{Chang} performed calculations using the pseudopotential method
    within the local density approximation. However, they found it necessary to apply a shift (of 2.3 mRy per P atom) to their calculated total energy of the $A$7 phase in order to reproduce
   the experimental value of the transition pressure, which is pointing out the inadequacy of the LDA to describe correctly the energetics of black phosphorus.
   This is in contradiction with the results of Ahuja\cite{Rajeev2003} who obtained a transition pressure of $P_t$= 4.5 GPa using the LMTO method and LDA.
 In the present work we also studied the  transition pressure    using Grimme's vdW functional. This leads to a value of $P_t= 0.8$ GPa,
 implying that although the Grimme functional is correctly ranking the $A$17 structure versus the $A$7 structure in terms of the total energy at $P=0$ GPa, the transition
 happens at a too low pressure in the calculation compared to the experimental pressure.
    Similar calculations were carried out using either of the PW91, PBE, OBS or TS functionals, but in all cases the transition pressure is found
    to be too low, within a 1 GPa range around the value obtained with the Grimme functional.

Possibly, a better agreement with the experimental  transition pressure could be obtained   using a more advanced description of  the van der Waals interactions,
 such as the adiabatic-connection fluctuation-dissipation theorem\cite{Gunnarsson, Langreth1975,Langreth1977} in the random phase approximation
 or quantum Monte-Carlo\cite{foulkes} method. However, these methods are numerically heavier than the simple dipole-dipole corrections
 considered here, and to our knowledge the implementation 
 of the corresponding forces necessary to perform the structural relaxation under pressure
   is not 
available.

\subsection{Optical Phonons}
The optical phonons of black phosphorus have been obtained by the linear response method within density functional perturbation theory. \cite{Gonze} 
In this method the force-constant matrix is obtained by differentiation of the Hellmann-Feynman forces on atoms with respect to the ionic coordinates.
In order to understand the optical phonons of black phosphorus, it is necessary to first consider the dynamical properties of this structure. The primitive cell contains four atoms leading to
12 vibrational modes, i. e. 9 optical and 3 acoustic phonons branches. From the group theoretical analysis of the C\emph{mca} space group,
the irreducible representation at the Brillouin zone center is
$$\Gamma_{acoustic} = B_{1u} + B_{2u} + B_{3u}$$ and
$$\Gamma_{optical} = B_{1u} + B_{2u} + A_u + 2A_g + B_{1g} + B_{2g} + 2B_{3g}$$

\par
\begin{figure}[h!]
\includegraphics[width=\linewidth,clip]{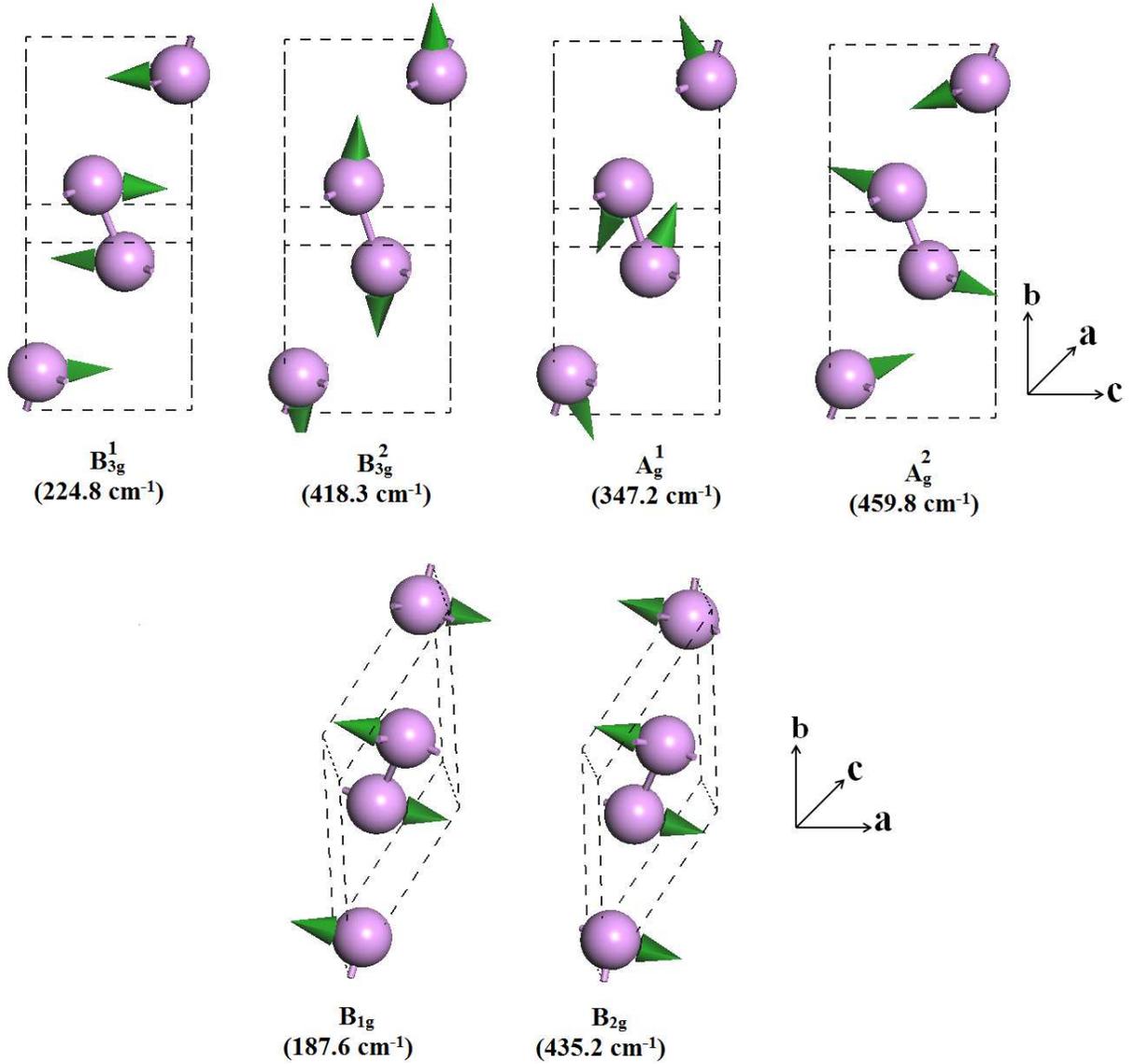}\\
\caption{(Color online) Snapshots of Raman active modes for primitive black phosphorus. The listed frequencies are as calculated within  GGA-Grimme at P=0.}
\label{Fig_modes}
 \end{figure}

\begin{figure}[h!]
\includegraphics[width=\linewidth,clip]{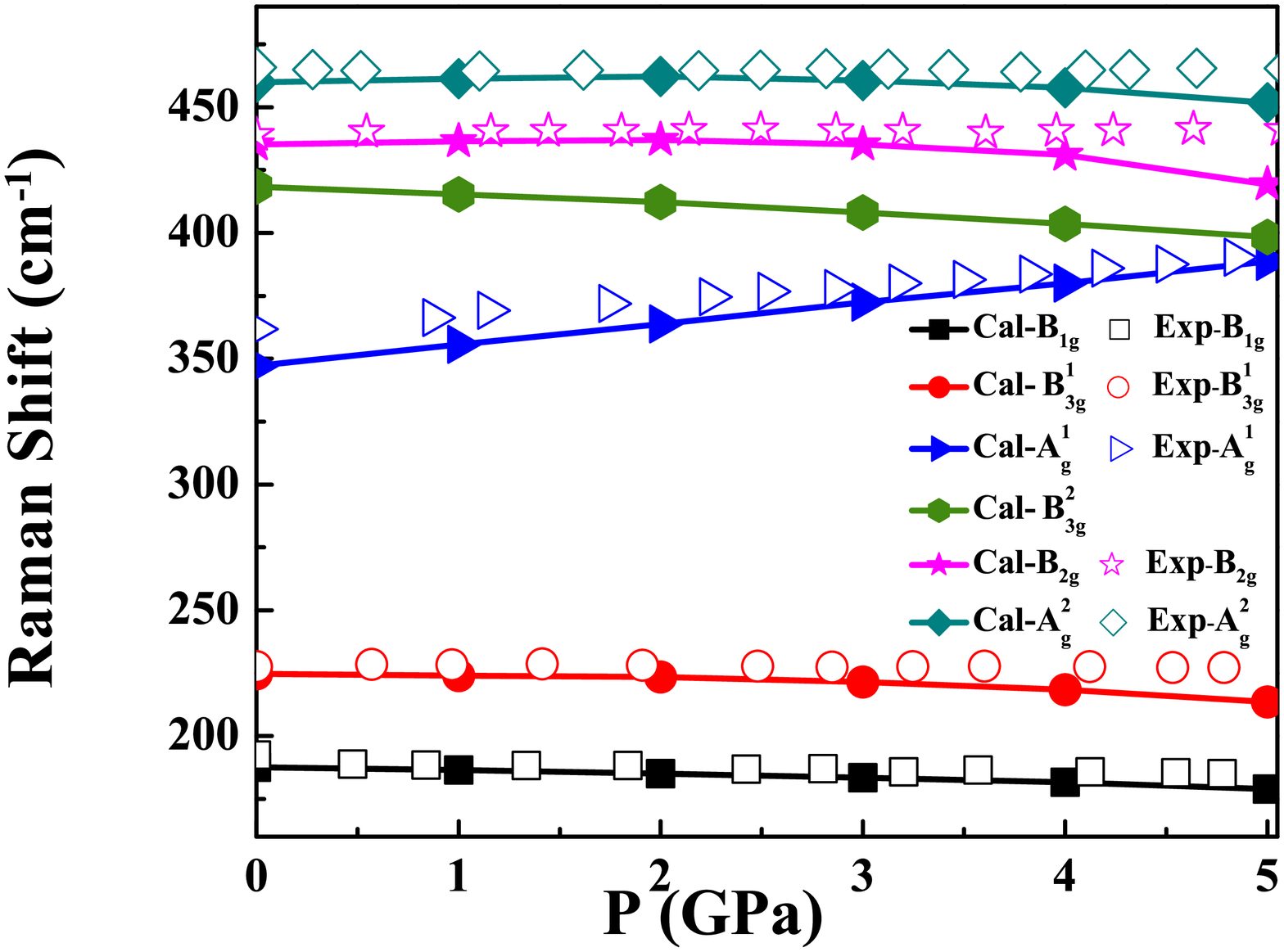}
\caption{(Colour online) Calculated (PBE-Grimme) Raman frequencies under pressure compared to experiments (Ref. \onlinecite{Kawamura}).
\label{Fig_Raman}
}
\end{figure}

 All the optical phonons are Raman active except the $B_{1u}$ and $B_{2u}$ modes, which are infrared active,  and the $A_u$ mode, which  is silent.
The optical phonons of black phosphorus were investigated by Sugai and Shirotani\cite{Sugai85} and by Akahama et al.\cite{Kawamura} The Raman frequencies turn out to be relatively insensitive to the details of the functional employed, for which reason we show the results obtained with the GGA-Grimme fuctional.
This is probably a reflection of the fact that the optical phonons are described by near-neighbour interatomic forces, which are dominated by covalent interactions, which are generally well accounted for by the LDA. The eigenvectors of the Raman modes, as calculated at ambient pressure within the GGA-Grimme approach are illustrated in Fig. \ref{Fig_modes}, in good agreement with earlier work.\cite{Sugai85,Kawamura}
From the calculated Raman eigen-modes, the $A_g^1$ , $A_g^2$ and $B_{3g}^2$ modes are seen to constitute inter-layer vibrations with a strong component along the $b$ direction.
In the modes $B_{1g}$ and $B_{2g}$ the atoms move along the $a$ direction, while in the $B_{3g}^1$ mode the atoms move along the $c$ direction.
Ref. \onlinecite{Kawamura} studied the pressure variation of the Raman shifts. In Fig. \ref{Fig_Raman} the calculated Raman frequencies under pressure up to 5 GPa are compared with these
experimental results.
 The agreement between theory and experiment is excellent. The strongest pressure dependence is seen for the inter-layer $A_g^1$ mode, for which
 the pressure coefficient is $d\omega/dP=8.2$ cm$^{-1}$/GPa in theory, and $d\omega/dP=5.78$ cm$^{-1}$/GPa according to experiment.\cite{Kawamura}

\section{CONCLUSIONS}

First-principles calculations based on the DFT have been used to investigate the structural properties of black phosphorus within LDA and GGA. It was found that
the calculated material parameters differ significantly from the experimental data, while including empirical methods to account for van der Waals interactions (DFT-D) leads to 
structural properties in good agreement with experiments.  This demonstrates that van der Waals interactions play a major role in black phosphorus. Among the OBS, TS and Grimme parametrizations of van der Waals interactions, the calculated structural parameters are in closest agreement with experimental data when  PBE-Grimme is used. Also,  the influence of external pressure on the lattice parameters were calculated.
The $b$ lattice parameter was found to be most sensitive to which vdW functional was used, reflecting the dominance of    van der Waals interactions   along the crystallographic $b$-direction.
Further,  the elastic properties of single crystal as well as polycrystalline black phosphorus were examined using the  GGA-PBE-Grimme functional.
Our calculations confirm that black phosphorus is a  mechanically stable anisotropic material, and   that the system is stiffest along the $a$-axis.
The calculated shear anisotropy factors for different directions indicate strong anisotropies for shears perpendicular to the $a$ and $b$ axes.
The polycrystalline bulk, shear and Young's  moduli, as well as  Poisson's ratio were calculated. From the ratio of bulk to shear modulus, black phosphorus is found to be brittle.
Finally, the frequencies of Raman active vibrational modes were calculated and they are in excellent agreement with the available experimental data.


\section{ACKNOWLEDGMENTS}
S. A. would like to thank DRDO through ACRHEM for financial support and thank CMSD, University of Hyderabad, for providing computational facilities.

*Corresponding author,
E-mail: gvsp@uohyd.ernet.in


\begin{thebibliography}{}

\bibitem{Boocker}
S. B\"ocker and M. H\"aser,  Z. Anorg. Allg. Chem. \textbf{321}, 258 (1995).

\bibitem{Morita}
A. Morita, Appl. Phys. A: Solid Surf. \textbf{39}, 227 (1986).

\bibitem{Akahama}
Y. Akahama, S. Endo and S. Narita, Physica \textbf{139-140B}, 397 (1986).

\bibitem{Cartz}
L. Cartz, S. R. Srinivasa, R. J. Riedner, J. D. Jorgensen and T. G. Worlton, J. Chem. Phys. \textbf{71}, 1718 (1979).

\bibitem{John}
J. C. Jamieson, Science \textbf{139}, 1291 (1963).

\bibitem{Duggin}
M. J. Duggin, J. Phys. Chem. Solids \textbf{33}, 1267 (1972).

\bibitem{Kikegawa}
T. Kikegawa and H. Iwasaki,  Acta Cryst. \textbf{B 39}, 158 (1983).

\bibitem{Suzuki}
T. Suzuki, T. Yagi, S. Akimito, T. Kawamura, S. Toyoda and  S. Endo, J. Appl. Phys. \textbf{54}, 748 (1983).

\bibitem{Okajima}
M. Okajima, S. Endo, Y. Akahama and S. Narita, J. Appl. Phys. \textbf{23}, 15 (1984).

\bibitem{Akahama137}
Y. Akahama, M. Kobayashi, and H. Kawamura, Phys. Rev. B \textbf{59}, 8520 (1999).

\bibitem{Akahama262}
Y. Akahama and H. Kawamura, S. Carlson, T. Le Bihan, and D. H\"ausermann, Phys. Rev. B \textbf{61}, 3139 (2000).

\bibitem{Maruyama}
H. Asahina, Y. Maruyama and A. Morita, Physica \textbf{117B - 118B}, 419 (1983).

\bibitem{Narita}
S. Narita, Y. Akahama, Y. Tsukiyama, K. Muro, S. Mori, S. Endo, M. Taniguchi, M. Seki, S. Suga, A. Mikuni and H. Kanzaki, Physica B+C, \textbf{117 - 118}, 422 (1983).

\bibitem{Takahashi}
T. Takahashi, K. Shirotani, S. Suzuki and T. Sagawa, Solid State Commun. \textbf{45}, 945 (1983).

\bibitem{Ikezawa}
M. Ikezawa, Y. Kondo and I. Shirotani, J. Phys. Soc. Japan \textbf{52}, 1505 (1983).

\bibitem{Fujii}
Y. Fujii, Y. Akahama, S. Endo, S. Narita, Y. Yamada and  G. Shirane, Solid State Commun. \textbf{44} 579, (1982).

\bibitem{Erik}
B. Nolang, O. Eriksson and B. Johansson, J. Phys. Chem. Solids \textbf{51}, 1025 (1990).

\bibitem{Rajagopal}
M. Rajagopalan, M. Alouani and N. E. Christensen, J. Low Temp. Phys. \textbf{75}, 1 (1989).

\bibitem{Rajeev2003}
R. Ahuja, Phys. Stat. Sol. B \textbf{235}, 282 (2003).

\bibitem{OKA}
O. K. Andersen, Phys. Rev. B  \textbf{12}, 3060 (1975).

\bibitem{Asahina}
H. Asahina and A. Morita, J. Phys. C: Solid State Phys. \textbf{17}, 1839 (1984).

\bibitem{Kaneta}
C. Kaneta, H. Katayama-Yoshida and A. Morita, J. Phys. Soc. Japan \textbf{55}, 1213 (1986).

\bibitem{Schiferl}
D. Schiferl, Phys. Rev. B \textbf{19}, 806 (1979).

\bibitem{Chang}
K. J. Chang and M. L. Cohen, Phys. Rev. B \textbf{33}, 6177 (1986).

\bibitem{Shirotani}
I. Shirotani, K. Tsuji, M. Imai, H. Kawamura, O. Shimomura, T. Kikegawa and T. Nakajima, Phys. Lett. A \textbf{144}, 102 (1990).

\bibitem{Kawamura}
Y. Akahama, M. Kobayashi and H. Kawamura, Solid State Commun. \textbf{104}, 311 (1997).

\bibitem{Shindo}
A. Nishikawa, K. Niizeki and K. Shindo, Phys. Status Solidi B \textbf{223}, 189 (2001).

\bibitem{Ehlers}
F. J. H. Ehlers and N. E. Christensen, Phys. Rev. B \textbf{69}, 214112 (2004).

\bibitem{Prytz}
\O. Prytz and E. Flage-Larsen, J. Phys. Cond. Matt. \textbf{22}, 015502 (2009).

\bibitem{Clark}
S. M. Clark and J. M. Zaug, Phys. Rev. B \textbf{82}, 134111 (2010).

\bibitem{Grin}
S. E. Boulfelfel, G. Seifert, Yu. Grin and S. Leoni, Phys. Rev. B \textbf{85}, 014110 (2012).

\bibitem{Yanlan}
Y. Du, C. Ouyang, S. Shi and M. Lei, J. Appl. Phys. \textbf{107}, 093718 (2010).

\bibitem{Grimme}
S. Grimme, J. Comp. Chem. \textbf{27}, 1787 (2006).
\bibitem{Payne}
M. C. Payne, M. P. Teter, D. C.  Allan, T. A. Arias and J. D. Joannopoulos, Rev. Mod. Phys. \textbf{64}, 1045 (1992).

\bibitem{Segall}
M. D. Segall, P. J. D. Lindan, M. J. Probert, C. J. Pickard, P. J. Hasnip, S. J. Clark and  M. C. Payne, J. Phys. Cond. Matt. \textbf{14}, 2717 (2002).

\bibitem{Vanderbilt}
D. Vanderbilt, Phys. Rev. B \textbf{41}, 7892 (1990).

\bibitem{Ceperley}
D. M. Ceperley and  B. J. Alder, Phys. Rev. Lett. \textbf{45},  566 (1980).

\bibitem{PPerdew}
J. P. Perdew and A. Zunger, Phys. Rev. B \textbf{23}, 5048 (1981).

\bibitem{Wang}
J. P. Perdew and Y. Wang, Phys. Rev. B \textbf{45}, 13244 (1992).

\bibitem{Perdew}
J. P. Perdew, K. Burke and M. Ernzerhof,  Phys. Rev. Lett. \textbf{77}, 3865 (1996).

\bibitem{OBS}
F. Ortmann, F. Bechstedt and W. G. Schmidt, Phys. Rev. B \textbf{73}, 205101 (2006).

\bibitem{TS}
A. Tkatchenko and M. Scheffler, Phys. Rev. Lett. \textbf{102}, 073005 (2009).

\bibitem{sebvasp} T. Bucko, J. Hafner, S. Leb\`egue, and J. G. Angy\'an, J. Phys. Chem. A \textbf{114}, 11814 (2010)
\bibitem{sebpccp} T. Bucko, J. Hafner, S. Leb\`egue, and J. G. Angy\'an, Phys. Chem. Chem. Phys. \textbf{14}, 5389 (2012)
\bibitem{obsapp1} T. F. Zhang, M. Sacchi, D. A. King, and S. M. Driver, J. Phys. Chem. C \textbf{116}, 5637 (2012)
\bibitem{ts1app} B. Santra, J. Klimes, D. Alfe, A. Tkatchenko, B. Slater, A. Michaelides, R. Car, and M. Scheffler, Phys. Rev. Lett \textbf{107}, 185701 (2011)
\bibitem{ts2app} N. Marom, J. Bernstein, J. Garel, A. Tkatchenko, E.Joselevich, L. Kronik, and O. Hod, Phys. Rev. Lett \textbf{105}, 046801 (2010)

\bibitem{Monkhorst}
H. J. Monkhorst and J. Pack,  Phys. Rev. B \textbf{13}, 5188 (1976).

\bibitem{Pulay}
P. Pulay, Mol. Phys. \textbf{17}, 197 (1969).

\bibitem{Mehl}
M. J. Mehl, J. E. Osburn, D. A. Papaconstantopoulus and B. M. Klein, Phys. Rev. B \textbf{41}, 10311 (1990).

\bibitem{Ravindran}
P. Ravindran, L. Fast, P. A. Korzhavyi and B. Johansson, J. Appl. Phys. \textbf{84}, 4891 (1998).

\bibitem{Hanayama}
Y. Kozuki, Y. Hanayama, M. Kimura, T. Nishitake and S. Endo, J. Phys. Soc. Japan \textbf{60}, 1612 (1991).

\bibitem{Masahito}
M. Yoshizawa, I. Shirotani and T. Fujimura, J. Phys. Soc. Japan \textbf{55}, 1196 {1986}.

\bibitem{HL}
L. Hedin and B. I. Lundqvist, J. Phys. C \textbf{4}, 2064 (1971).

\bibitem{Akai}
T. Akai, S. Endo, Y. Akahama, K. Koto, Y. Maruyama, High Pressure Research  \textbf{1}, 115 (1989).

\bibitem{Born}
M. Born and  K. Huang, {\it Dynamical Theory of Crystal Lattices}, Oxford University Press, (Oxford, 1998).

\bibitem{Voigt}
W. Voigt, Ann. Phys. (Leipzig) \textbf{38}, 573 (1889).

\bibitem{Reuss}
A. Reuss, Z. Angew. Math. Phys. \textbf{9}, 49 (1929).

\bibitem{Hill}
R. Hill, Proc. Phys. Soc. London \textbf{65}, 350 (1952).

\bibitem{Pugh}
S. F. Pugh, Philos. Mag. \textbf{45}, 823 (1954).

\bibitem{bheem}
Ch. Bheema Lingam, K. Ramesh Babu, Surya. P. Tewari and G. Vaitheeswaran, J. Comp. Chem. \textbf{32}, 1734 (2011).

\bibitem{Burdett}
J. K. Burdett and S. Lee, J. Solid State Chem.  \textbf{44}, 415 (1982).

\bibitem{Sugai85}
S. Sugai and I. Shirotani, Solid State Commun. \textbf{53}, 753 (1985).

\bibitem{Gunnarsson}
O. Gunnarsson and B. I. Lundqvist, Phys. Rev. B \textbf{13}, 4274 (1976).

\bibitem{Langreth1975}
D. C. Langreth and J. P. Perdew, Solid State Comm. \textbf{17}, 1425 (1975).

\bibitem{Langreth1977}
D. C. Langreth and J. P. Perdew, Phys. Rev. B \textbf{15}, 2884 (1977).

\bibitem{foulkes}
W. M. C. Foulkes, L. Mitas, R. J. Needs, and G. Rajagopal , Rev. Mod. Phys. \textbf{73}, 33 (2001).

\bibitem{Gonze}
X. Gonze, Phys. Rev. B \textbf{55}, 10337 (1977).


\end{thebibliography}
\end{document}